\documentclass[draft,eqsecnum,nofootinbib,aps]{revtex4}
\renewcommand{\theequation}{\arabic{equation}}
\def\beq{\begin{equation}}
\def\eeq{\end{equation}}
\def\bea{\begin{eqnarray}}
\def\eea{\end{eqnarray}}
\def\nn{\nonumber}

\begin{document}
\title{Warped products and black holes}
\author{Soon-Tae Hong}
\email{soonhong@ewha.ac.kr} 
\affiliation{Department of Science Education, Ewha Womans University, Seoul 120-750 Korea}
\date{August 1, 2003}
\begin{abstract}
We apply the warped product spacetime scheme to the  Banados-Teitelboim-Zanelli 
black holes and the Reissner-Nordstr\"om-anti-de Sitter black hole to 
investigate their interior solutions in terms of warped products.  
It is shown that there exist no discontinuities of the Ricci and Einstein curvatures 
across event horizons of these black holes.

\end{abstract}
 \maketitle

\section{Introduction}
\setcounter{equation}{0}
\renewcommand{\theequation}{\arabic{section}.\arabic{equation}}

A warped product spacetime was introduced by Bishop and O'Neill long ago~\cite{bo}.  
The warped product scheme was later applied to general 
relativity~\cite{be96} and semi-Riemannian geometry~\cite{be79}.  Recently, this warped products 
were extended to multiply warped products with non-smooth metric~\cite{choi00}, and 
the Banados-Teitelboim-Zanelli (BTZ) black hole~\cite{hong03grg} and Reissner-Nordstr\"om-anti-de Sitter (RN-AdS) black hole~\cite{hong03math}. 

On the other hand, the concept of the warped products was used in higher dimensional theories.  
For instance, the warped products were exploited in the Randall-Sundrum model in five-dimensions~\cite{rs1,rs2} 
and in the Kaluza-Klein supergravity theory in seven-dimensions~\cite{kksug}.

In this paper we will briefly analyze the multiply warped product manifold associated
with the (2+1) BTZ black holes and (3+1) RN-AdS metric to investigate the physical
properties inside the event horizons.  In Section 2 we will introduce the concepts of 
multiply warped product manifolds. We will then apply this warped product scheme
to the (2+1) BTZ black holes in Section 3, and to the (3+1) RN-AdS black hole in Section 4 so that 
we can explicitly obtain the Ricci and Einstein curvatures inside the event horizons of 
these black holes.

\section{ Warped product spacetime}
\setcounter{equation}{0}
\renewcommand{\theequation}{\arabic{section}.\arabic{equation}}

Mathematically, a multiply warped product manifold is defined as 
$(M=B\times F_1\times...\times F_{n}, g)$ 
consisting of the Riemannian base manifold $(B, g_B)$ and fibers $(F_i,g_i)$ ($i=1,...,n$) associated with 
the Lorentzian metric,
\beq
g=\pi_{B}^{*}g_{B}+\sum_{i=1}^{n}(f_{i}\circ\pi_{B})^{2}\pi_{i}^{*}g_{i},
\eeq
where $\pi_B$, $\pi_{i}$ are natural projections of $B\times F_1\times...\times F_n$
onto $B$ and $F_{i}$, respectively, and $f_{i}$ are positive warping functions.

For a specific case of $(B=R,g_B=-d\mu^{2})$, the Lorentzian metric is of the form  
\beq
g=-d\mu^{2}+\sum_{i=1}^{n}f_{i}^{2}g_{i}.
\eeq
For instance, the Randall-Sundrum model metric can be regarded as the warped product manifold with 
the metric
\beq
g=-N^{2}(t,y)dt^{2}+A^{2}(t,y)d\vec{x}^{2}+dy^{2}.
\eeq

Next, we consider the warped products with single discontinuity.  Let 
$M=M_{0}\times_{f_1} F_1\times\cdots\times_{f_n} F_n$ be multiply warped products with 
Riemannian curvature tensor $R$. If $X$, $Y\in V(M_0)$, $U_i$, $V_i\in V(F_i)$ ($i=1,2,...,n$),
$d_i={\rm dim}~F_i$, $f_i\in C^0(S)$ at a single point $p\in
M_{0}$ and $S=\{p\}\times _{f_1} F_1\times\cdots\times_{f_n} F_n$, we then obtain the Ricci components 
of the form
\begin{eqnarray} 
{\rm Ric}(X,Y)&=&-\sum_{i=1}^{n}d_i
X^{1}Y^{1}\frac{f_i''+\delta(p)
\left({f_i'}^{+}-{f_i'}^{-}\right)}{f_i},\nonumber\\
{\rm Ric}(X,U_i)&=&0,\nonumber\\
{\rm Ric}(U_i,V_i)&=&{}^{F_{i}}{\rm Ric}(U_i, V_i)+\langle U_i, V_i\rangle
\frac{f_i''+\delta(p)({f_i'}^{+}-{f_i'}^{-})}{f_i}
\nn\\
&&+\langle U_i, V_i\rangle\delta(p)\left[
(d_{i}-1)\frac{{f_{i}'}^{+}-{f_{i}'}^{-}}{f_{i}^{2}}
+\sum_{j\neq i} d_{j}\frac{\langle {f_i'}^{+}-{f_i'}^{-},\ {f_j'}^{+}
-{f_j'}^{-}\rangle}{f_i f_j}\right],\nonumber\\
{\rm Ric}(U_i, U_j)&=&0~~~ {\rm for}~~i\neq j,
\eea
where $X=X^{1}\partial/\partial_{t}$ and $Y=Y^{1}\partial/\partial_{t}$.  Here note that we have 
discontinuity contributions associated with $\delta(p)$
at the point $p$ since we assume $f_{i}$ are $C^{0}$-functions.

\section{BTZ black holes in warped products}
\setcounter{equation}{0}
\renewcommand{\theequation}{\arabic{section}.\arabic{equation}}

In this section, we apply the warped product scheme to the BTZ metric to investigate the inner 
structure of the black hole.  We start with the static BTZ three-metric of the form 
\beq
ds^{2}=N^{2}dt^{2}-N^{-2}dr^{2}+r^{2}d\phi^{2}.
\eeq
Here the lapse function for interior solution is given by 
\beq
N^{2}=m-\frac{r^{2}}{l^{2}},
\eeq
which can be rewritten in terms of the event horizon $r_{H}=m^{1/2}l$ in the region $r<r_{H}$ 
as follows
\beq
N^{2}=\frac{(r_{H}+r)(r_{H}-r)}{l^{2}}.
\eeq
As you see, the lapse function is positive in the region $r<r_{H}$ and thus this lapse function is 
well-defined inside the event horizon.  So, the BTZ three-metric has a signature $(+,-,+,+)$ instead of 
$(-,+,+,+)$.

Next we define a new coordinate $\mu$ as 
\beq
d\mu^{2}=N^{-2}dr^{2}.
\label{mu}
\eeq
From this definition, $\mu$ can be given by a integration 
\beq
\mu=\int_{0}^{r}dx~\frac{l}{[(r_{H}+x)(r_{H}-x)]^{1/2}},
\eeq
whose analytic solution is simply written as follows
\beq
\mu=l\sin^{-1}\left(\frac{r}{r_{H}}\right)=F(r).
\eeq
The function $F(r)$ then satisfies boundary conditions
\beq
{\rm lim}_{r\rightarrow r_{H}}F(r)=\frac{l\pi}{2},~~~
{\rm lim}_{r\rightarrow 0}F(r)=0,
\eeq
and $dr/d\mu >0$ implies $F^{-1}$ is well-defined function.

Using the coordinate $\mu$, we rewrite the BTZ metric in terms of the warped products 
\beq
ds^{2}=-d\mu^{2}+f_{1}(\mu)^{2}dt^{2}+f_{2}^{2}(\mu)d\phi^{2},
\label{dsmu}
\eeq
where $f_{1}$ and $f_{2}$ are warping functions given in terms of $\mu$ as follows
\beq
f_{1}(\mu)=\left(m-\frac{F^{-2}(\mu)}{l^{2}}\right)^{1/2},~~~
f_{2}(\mu)=F^{-1}(\mu).
\label{warpf}
\eeq
After some algebra using the warped product metric in (\ref{dsmu}), we can obtain the 
nonvanishing Ricci curvature components
\bea
R_{\mu\mu}&=&-\frac{f_{1}^{\prime\prime}}{f_{1}}-\frac{f_{2}^{\prime\prime}}{f_{2}},
\nonumber\\
R_{tt}&=&\frac{f_{1}f_{1}^{\prime}f_{2}^{\prime}}{f_{2}}+f_{1}f_{1}^{\prime\prime},
\nonumber\\
R_{\phi\phi}&=&\frac{f_{1}^{\prime}f_{2}f_{2}^{\prime}}{f_{1}}+f_{2}f_{2}^{\prime\prime}.
\eea

Exploiting the explicit expressions for $f_{1}$ and $f_{2}$ in (\ref{warpf}), we can 
obtain the identities for $f_{1}$, $f_{1}^{\prime}$ and $f_{1}^{\prime\prime}$ in terms 
of $f_{1}$, $f_{2}$ and their derivatives as below
\bea 
f_{1}&=&f_{2}^{\prime},\nonumber\\
f_{1}^{\prime}&=&-\frac{f_{2}}{l^{2}},\nonumber\\
f_{1}^{\prime\prime}&=&\frac{f_{1}f_{1}^{\prime}}{f_{2}},
\eea
so that inside event horizons we can obtain the Ricci curvatures of the simple form
\bea
R_{\mu\mu}&=&-\frac{2f_{1}^{\prime}}{f_{2}},\nonumber\\
R_{tt}&=&\frac{2f_{1}^{2}f_{1}^{\prime}}{f_{2}},\nonumber\\
R_{\phi\phi}&=&2f_{2}f_{1}^{\prime}.
\label{btzricci}
\eea
Similarly, we can also evaluate the Einstein scalar curvature inside the event horizon as follows 
\beq
R=-\frac{6}{l^{2}}.
\eeq

Next, we investigate the relations between the inner solutions above and the well-known exterior 
solutions.  As you see, outside event horizon $r_{H}$, the BTZ three-metric is given by
\beq
ds^{2}=-(-m+\frac{r^{2}}{l^{2}})^{2}dt^{2}
+(-m+\frac{r^{2}}{l^{2}})^{-2}dr^{2}+r^{2}d\phi^{2}.
\eeq
Using this exterior metric, we can obtain the Ricci and Einstein curvatures explicitly in 
terms of the warping functions $f_{i}$ defined in (\ref{warpf}),
\bea
R_{rr}&=&-\frac{2f_{1}^{\prime}}{f_{1}^{2}f_{2}},\nonumber\\
R_{tt}&=&\frac{2f_{1}^{2}f_{1}^{\prime}}{f_{2}},\nonumber\\
R_{\phi\phi}&=&2f_{2}f_{1}^{\prime},\nonumber\\
R&=&-\frac{6}{l^{2}}.
\label{btzricci2}
\eea
Here note that the Einstein scalar curvature $R$ is identical to 
that of the interior case and the Ricci components $R_{tt}$ and $R_{\phi\phi}$ are also the same 
as those of interior case.  Moreover, from the definition of the coordinate $\mu$ in (\ref{mu}), 
we can find the following identity 
\beq
R_{\mu\mu}=f_{1}^{2}R_{rr}
\eeq
which is also attainable by comparing $R_{\mu\mu}$ and $R_{rr}$ components in (\ref{btzricci}) 
and (\ref{btzricci2}).  We can thus conclude that all the Ricci
components and Einstein scalar curvature have identical forms both in
exterior and interior of event horizon $r_{H}$ without discontinuities.

Next, we consider the charged BTZ black hole whose lapse function for interior solution is 
given by  
\beq
N^{2}=m-\frac{r^{2}}{l^{2}}+2Q^{2}\ln~r,
\eeq
where we have an additional term proportional to $Q^{2}$ where $Q$ is the charge of the BTZ 
black hole.  Similar to the static BTZ case, we can find the Ricci and Einstein curvatures 
inside event horizon $r_{H}$ as follows 
\bea
R_{\mu\mu}&=&-\frac{2f_{1}^{\prime}}{f_{2}}+\frac{2Q^{2}}{f_{2}^{2}},\nonumber\\
R_{tt}&=&\frac{2f_{1}^{2}f_{1}^{\prime}}{f_{2}}-\frac{2Q^{2}f_{1}^{2}}{f_{2}^{2}},\nonumber\\
R_{\phi\phi}&=&2f_{2}f_{1}^{\prime},\nonumber\\
R&=&-\frac{6}{l^{2}}+\frac{2Q^{2}}{f_{2}^{2}}. 
\eea
Note that we have charge contributions in the Ricci components $R_{\mu\mu}$ and $R_{tt}$ and 
also in the Einstein curvature $R$.

\section{RN-AdS black hole in warped products}
\setcounter{equation}{0}
\renewcommand{\theequation}{\arabic{section}.\arabic{equation}}

Next, in this section we consider the RN-AdS metric in the warped product scheme to study the inner 
structure of the black hole.  The (3+1) RN-AdS four-metric is given by 
\beq
ds^{2}=N^{2}dt^{2}-N^{-2}dr^{2}+r^{2}(d\theta^{2}+\sin^{2}\theta d\phi^{2}),
\eeq
whose lapse function in region between event horizons $r_{-}$ and $r_{+}$ is described as
\beq
N^{2}=-1+\frac{2m}{r}-\frac{Q^{2}}{r^{2}}-\frac{r^{2}}{l^{2}},
\eeq
with the charge $Q$ and the cosmological constant $1/l^{2}$.  Again note that the four-metric has a positive signature in time direction as mentioned in the BTZ case.  If we define the event 
horizons $r_{-}$ and $r_{+}$ by solving the equation 
\beq
0=-1+\frac{2m}{r_{\pm}}-\frac{Q^{2}}{r_{\pm}^{2}}-\frac{r_{\pm}^{2}}{l^{2}},
\eeq
we can then rewrite $N^{2}$, $Q^{2}$ and $l^{2}$ in terms of $r_{-}$ and $r_{+}$ as follows
\bea
N^{2}&=&\frac{(r_{+}-r)(r-r_{-})}{r^{2}l^{2}}\left(r^{2}+(r_{+}+r_{-})r+\frac{Q^{2}l^{2}}
{r_{+}r_{-}}\right),\nn\\
Q^{2}&=&\frac{r_{+}r_{-}[2m(r_{+}^{2}+r_{+}r_{-}+r_{-}^{2})-r_{+}r_{-}(r_{+}+r_{-})]}
{(r_{+}+r_{-})(r_{+}^{2}+r_{-}^{2})},\nonumber\\
l^{2}&=&\frac{(r_{+}+r_{-})(r_{+}^{2}+r_{-}^{2})}{2m-r_{+}-r_{-}}.
\end{eqnarray}

As in the BTZ black hole case, we define a new coordinate $\mu$ as
\beq
d\mu^{2}=N^{-2}dr^{2},
\label{mu2}
\eeq
which is integrated to yield the following somewhat complicated expression
\beq
\mu=\int_{r_{-}}^{r}\frac{dx~xl}{[(r_{+}-x)(x-r_{-})
(x^{2}+(r_{+}+r_{-})x+Q^{2}l^{2}/r_{+}r_{-})]^{1/2}}=F(r).
\label{rnmu}
\eeq
Here note that $dr/d\mu >0$ implies $F^{-1}$ is well-defined function.

Exploiting the coordinate $\mu$ defined in (\ref{rnmu}), we now rewrite the RN-AdS 
black hole metric in terms of the warped products
\beq
ds^{2}=-d\mu^{2}+f_{1}^{2}(\mu)d\nu^{2}+f_{2}^{2}(\mu)(d\theta^{2}+\sin^{2}\theta d\phi^{2}),
\label{rnmetricf}
\eeq
where the warping functions $f_{1}$ and $f_{2}$ are given by 
\begin{eqnarray}
f_{1}(\mu)&=&\left(-1+\frac{2m}{F^{-1}(\mu)}-\frac{Q^{2}}
{F^{-2}(\mu)}-\frac{F^{-2}(\mu)}{l^{2}}\right)^{1/2},\nonumber\\
f_{2}(\mu)&=&F^{-1}(\mu).
\end{eqnarray}

Again, after some tedious algebra using the metric (\ref{rnmetricf}), we can obtain 
the Ricci curvature components in terms of the warping functions
\begin{eqnarray}
R_{\mu\mu}&=&-\frac{f_{1}^{''}}{f_{1}}-\frac{2f_{2}^{''}}{f_{2}},
\nonumber\\
R_{tt}&=&\frac{2f_{1}f_{1}^{'}f_{2}^{'}}{f_{2}}+f_{1}f_{1}^{''},
\nonumber\\
R_{\theta\theta}&=&\frac{f_{1}^{'}f_{2}f_{2}^{'}}{f_{1}}+f_{2}f_{2}^{''}
+f_{2}^{{'}^2}+1,
\nonumber\\
R_{\phi\phi}&=&\left(\frac{f_{1}^{'}f_{2}f_{2}^{'}}{f_{1}}+f_{2}f_{2}^{''}
+f_{2}^{{'}^2}+1 \right)\sin^{2}\theta.
\end{eqnarray}

As in the case of the BTZ black hole, we can find the following identities for $f_{1}$, 
$f_{1}^{\prime}$ and $f_{1}^{\prime\prime}$ 
\begin{eqnarray}
f_{1}&=&f_{2}^{'},\nonumber\\
f_{1}^{'}&=&-\frac{m}{f_{2}^{2}}+\frac{Q^{2}}{f_{2}^{3}}-\frac{f_{2}}{l^{2}},\nonumber\\
f_{1}^{''}&=&-\frac{2f_{1}f_{1}^{'}}{f_{2}}-\frac{Q^{2}f_{1}}{f_{2}^{4}}-\frac{3f_{1}}{l^{2}},
\end{eqnarray}
which can be used to evaluate the Ricci curvatures in the region between $r_{-}$ and $r_{+}$
\begin{eqnarray}
R_{\mu\mu}&=&\frac{Q^{2}}{f_{2}^{4}}+\frac{3}{l^{2}},\nonumber\\
R_{tt}&=&-\frac{Q^{2}f_{1}^{2}}{f_{2}^{4}}-\frac{3f_{1}^{2}}{l^{2}},\nonumber\\
R_{\theta\theta}&=&\frac{Q^{2}}{f_{2}^{2}}-\frac{3f_{2}^{2}}{l^{2}},\nonumber\\
R_{\phi\phi}&=&\left(\frac{Q^{2}}{f_{2}^{2}}-\frac{3f_{2}^{2}}{l^{2}}\right)\sin^{2}\theta,
\label{rnricci}
\end{eqnarray}
to yield the Einstein curvature
\beq
R=-\frac{12}{l^{2}}.
\eeq

On the other hand, outside event horizon $r_{+}$, we have the RN-AdS four-metric of the 
well-known form 
\beq
ds^{2}=-(1-\frac{2m}{r}+\frac{Q^{2}}{r^{2}}+\frac{r^{2}}{l^{2}})^{2}dt^{2}
+(1-\frac{2m}{r}+\frac{Q^{2}}{r^{2}}+\frac{r^{2}}{l^{2}})^{-2}dr^{2}
+r^{2}(d\theta^{2}+\sin^{2}\theta d\phi^{2}),
\eeq
to yield the Ricci and Einstein curvatures outside event horizons in terms of the warping functions
$f_{1}$ and $f_{2}$
\begin{eqnarray}
R_{rr}&=&\frac{Q^{2}}{f_{1}^{2}f_{2}^{4}}+\frac{3}{f_{1}^{2}l^{2}},\nonumber\\
R_{tt}&=&-\frac{Q^{2}f_{1}^{2}}{f_{2}^{4}}-\frac{3f_{1}^{2}}{l^{2}},\nonumber\\
R_{\theta\theta}&=&\frac{Q^{2}}{f_{2}^{2}}-\frac{3f_{2}^{2}}{l^{2}},\nonumber\\
R_{\phi\phi}&=&\left(\frac{Q^{2}}{f_{2}^{2}}-\frac{3f_{2}^{2}}{l^{2}}\right)\sin^{2}\theta,\nonumber\\
R&=&-\frac{12}{l^{2}}.
\label{rnricci2}
\end{eqnarray}
Note that the Einstein scalar curvature $R$ and the Ricci components $R_{tt}$, 
$R_{\theta\theta}$ and $R_{\phi\phi}$ are equal to those in the region between $r_{-}$ and $r_{+}$.  
Moreover, using the definition of the coordinate $\mu$ in (\ref{mu2}), we can 
find the following identity 
\beq
R_{\mu\mu}=f_{1}^{2}R_{rr},
\eeq
which is also attainable from the Ricci components $R_{\mu\mu}$ and $R_{rr}$ in (\ref{rnricci}) 
and (\ref{rnricci2}).  As in the BTZ case, all the Ricci components and Einstein scalar curvature 
thus have identical forms both in exterior and interior of event horizon $r_{+}$ without 
discontinuities.

\section{Conclusions}
\setcounter{equation}{0}
\renewcommand{\theequation}{\arabic{section}.\arabic{equation}}

In conclusion, we have studied the warped product spacetime to obtain the interior 
solutions of the BTZ black holes and the RN-AdS black hole in terms of warping functions.  
In both cases, there exist no discontinuities of the Ricci and Einstein curvatures 
across event horizons of these black holes.

\acknowledgments The author would like to thank Itzhak Bars for initial discussions. This
work is supported in part by the Korea Science and Engineering Foundation Grant
R01-2000-00015.

\end{document}